# DEEP LEARNING FOR MUSIC GENERATION. FOUR APPROACHES AND THEIR COMPARATIVE EVALUATION

Razvan PAROIU[1], Stefan TRAUSAN-MATU[2]

*This paper introduces four different artificial intelligence algorithms for music generation and aims to compare these methods not only based on the aesthetic quality of the generated music but also on their suitability for specific applications. The first set of melodies is produced by a slightly modified visual transformer neural network that is used as a language model. The second set of melodies is generated by combining chat sonification with a classic transformer neural network (the same method of music generation is presented in a previous research), the third set of melodies is generated by combining the Schillinger rhythm theory together with a classic transformer neural network, and the fourth set of melodies is generated using GPT3 transformer provided by OpenAI. A comparative analysis is performed on the melodies generated by these approaches and the results indicate that significant differences can be observed between them and regarding the aesthetic value of them, GPT3 produced the most pleasing melodies, and the newly introduced Schillinger method proved to generate better sounding music than previous sonification methods.*

## 1. Introduction

The music generated through artificial intelligence applications, while potentially lacking the emotional depth and connection attributed to human compositions, offers a significant advantage in terms of rapid production compared to music created by humans. This efficiency makes AI-generated music an asset for various creative projects in the same way text is used in the generation of low-cost advertising text [1]. Anyway, computer-generated music has found its way into public concerts [2, 3], where innovative techniques such as sonification are utilized to reduce the sense of artificiality for the audience [3].

The primary goal of this paper is to conduct a comparative analysis of twelve melodies generated by four distinct artificial intelligence algorithms utilized for music generation. These methods include a variety of techniques, such as sonification, music generation employing trained transformers, or fine-tuned large language models. Through this analysis, the paper aims to shed light on the diverse capabilities and characteristics of these AI-based approaches in generating musical content. Each of these techniques utilizes state-of-the-art Transformer neural networks. We selected this architecture not only for its exceptional performance in training on large datasets but also to maintain consistency across our methods. By employing the same neural network architecture, our evaluation focuses on highlighting the distinctiveness of each method rather than comparing different neural network structures.

---

[1] PhD student, Faculty of Automatics and Computer Science, University POLITEHNICA of Bucharest, Romania, e-mail: razvan.paroiu@gmail.com

[2] Prof., Faculty of Automatics and Computer Science, University POLITEHNICA of Bucharest, Romanian Academy Research Institute for Artificial Intelligence, Academy of Romanian Scientists, Romania, e-mail: stefan.trausan@upb.ro

This paper also introduces two novel techniques of generating music: the first is a unique transformer architecture (utilized for generating the first set of melodies) and the second is a hybrid method that combines Schillinger's theory of rhythm with neural networks (employed for generating the third set of melodies).

The paper continues with a section that presents related work. The methods used for music generation are introduced in the third section, which is followed by evaluation and conclusion sections.

## 2. Related work

### 2.1. Transformer neural networks in music

Music is characterized by a sequential dimension in time. Therefore, recurrent neural networks (RNN) such as long short-term memory (LSTM) or gated Recurrent Unit (GRU) [4] were a suitable solution for music generation. However, RNNs have some limitations. Transformers [5] are neural networks that make use of the attention mechanism [6] instead of relying on recurrent traversing of data, one major advantage being the parallelization, which is not available for RNNs. The transformer model has as encoder-decoder structure. The encoder component uses the attention mechanism to selectively attend distinct sections of the input, and the resulted encodings are then passed to the decoder. Transformer neural networks represent the most recent research in the field of artificial neural networks and have been frequently used to generate music.

The original transformer's [5] weakness is that it depends on absolute timing signals to encode the melody's note positions. This makes it difficult for the neural network to maintain track of regularity and periodicity which are frequently found in musical rhythm. A more modern approach that reduced this problem is the invention of music transformer [7] which relies on relative attention [8]. Relative attention is a mechanism that adjusts the attention of the neural network based on how far away two tokens are from one another and enables the model to focus more on relational aspects of music. The music transformer is able to produce music more consistently for a longer period of time than the original transformer [7], which loses the quality of the generated music after the initial sequence of notes. Other approaches that are also capable of learning long samples of music are based on transformers with reduced memory complexity [9].

Even though it was developed for vision related research, where it eliminated the usage of convolutional neural networks, the vision transformer [10] can also be applied to music with multiple voices, such as polyphonic music. In the original vision transformer, 2D pictures that have been flattened into data patches serve as the input data, but in the case of music, the neural network may be trained using the notes and durations from different voices rather than 2D graphics. In recent research, the vision transformer was successfully used for instrument recognition in polyphonic music [11].

In recent years, there was a lot of work in the domain of pre-trained neural networks. Because the transformer can generate output much faster than recurrent neural networks through the use of parallelized attention mechanism, a lot of pre-trained models, including BERT [12] and GPT3 [13], were trained and made available. The large quantity of data on which these models have been trained on, made them ideal to use in conjunction with other neural networks, not only in NLP but also in other tasks that can be related to text.

### 2.2. Music generation through sonification

RNNs and transformers generate music as the result of training with large datasets. Another method that may be used to produce music is sonification, which use of non-audio data to perceptualize it into the form of sound. It has been used to "hear the sound of magnetic fields" of different planets from our solar system that have been recorded and transmitted to earth by Voyager 1 and Voyager 2 satellites [14] or "the sound of gravitational waves" released by merging black holes from space detected by LIGO [15]. Some devices use sonification to transform information in an audible form, either of danger or alert like the Geiger counter that determines the level of radiation or of physiological data in the medical systems.

Sonification was also used to generate music from chat conversations between humans. According to the polyphonic model theory [16], successful human collaboration in conversations may be attributed to the same polyphonic counterpoint rules that are used to coordinate several voices performing at the same time. The polyphonic model originates from Mikhail Bakhtin's philosophy, which emphasized, that many characters from Fyodor Dostoevsky's writings follow a singular theme, which is similar to what is found in polyphonic music, and the conflicts that emerge between the characters seem to follow counterpoint rules [17]. Experiments that tested the polyphonic model theory through sonification were conducted [18]. Additional experiments have employed neural networks for sonification [19], and produced more diverse pitches than earlier methods, but further extended experiments were also needed.

### 2.3. The Schillinger theory of rhythm

The Schillinger theory of rhythm is part of a system of musical composition created by Joseph Schillinger. The system represents a method of musical composition that relies on mathematical principles but also allows the human composer to enhance the composition with emotions. Schillinger also applied mathematical reasoning in other arts, not just in music[3]. The system was used by many famous composers such as George Gershwin, Earle Brown, Benny Goodman and Glenn Miller.

Some examples of techniques that Schillinger used in the creation of musical rhythm are fractioning, grouping by pairs, coordination of time structures, or permutations applied to musical parameters [20]. In this paper, the musical rhythm was generated using permutations applied to musical parameters, which is a technique that obtains a temporal continuity by permuting parts of a melody inside a single voice (of an instrument). If the musical composition contains multiple voices, then there will also be a vertical distribution over multiple parts [20]. One example of a rhythm generated by Schillinger's method of permutations applied to musical parameters is presented in Figure 1. If a quarter note is labeled with A, an eighth note is labeled with B, two consecutive 16th notes are labeled with C, and one half note is labeled with D, then the two voices from Figure 1 become:

A B C D | D C B A
C D B A | C D B A

After the first two measures from the first voice, the next measures will repeat the same rhythm (A B C D | D C B A), while for the second voice the rhythm is composed only by the repetition of the first measure (C D B A). Furthermore, the song's melodic line (the pitches of notes) is entirely independent with the rhythm.

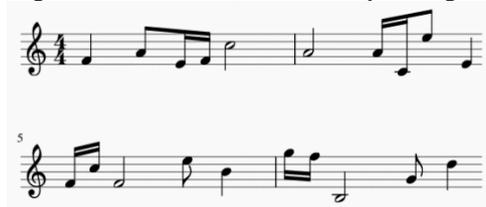

Fig. 1 - Rhythm created by Schillinger's theory of rhythm

### 3. The current methods chosen for music generation

The corpus used for training and testing all the neural networks from this paper is Lakh Midi Dataset (LMD) [21] which contains approximately 175.000 melodies from different genres [22] extracted from publicly accessible online sources. The melodies from the corpus are stored in MIDI format and for preprocessing was used the music21 python framework. Another popular music format is MusicXML, where the melodies are encoded in human readable xml notation. A disadvantage of MusicXML is its relatively large text size. An alternative is the ABC music format, which has a considerably smaller encoded text size than MusicXML and it is also readable by humans.

---

[3] Joseph Schillinger, Wikipedia, Mar. 17, 2023. https://en.wikipedia.org/wiki/Joseph_Schillinger

### 3.1. The first set of melodies

For generating the first set of melodies, a vision transformer [10] neural network was trained as a language model in order to produce both pitches and durations for the generated melodies. To avoid accumulating too much data in the RAM memory, the melodies from the corpus are fed into the neural network in batches of 100 melodies. After that, each batch is partitioned into 128 consecutive pitch sequences and 128 consecutive duration sequences. Each sequence consists of 150 consecutive notes extracted from 3 voices (50 consecutive notes from each voice). For this reason, the neural network is capable of producing music with only 3 voices. The transformer neural network also has the potential to be trained with larger batch sizes than traditional RNN due to the training process parallelization capabilities that are considerably improved.

Each pitch and duration is first encoded into distinct integers before being sent as input into the neural network. The pitch integer encoding takes up 126 different values (18 scales with 7 pitches each), whereas the durations encoding takes up 833 different values (the durations vocabulary of the entire corpus). The tokens from the input sequences are then embedded into tensors of 256 in size. By using the embedding size of 256, the neural network size will consist of 3.5 million parameters. The token embeddings are then combined (using addition) with the positional embeddings as shown in Figure 2. The transformer architecture encodes the position of each token from the sequence in order to recognize patterns between successive notes (patterns that produce the melody) and successive durations (patterns that form the rhythm). This is required to eliminate the LSTM and GRU neural networks recursiveness process that cannot be parallelized.

The sine and cosine functions are used for positional encoding. The following formulas, taken from the original publication of the transformer architecture [5], are used to generate the positions of the tokens:

$$PE_{(pos,2i)} = sin\big(pos \,/\, 10000^{2i/d_{model}}\big)$$
$$PE_{(pos,2i+1)} = cos\big(pos \,/\, 10000^{2i/d_{model}}\big)$$

where pos represents the position of the token in the sequence, 2i and 2i+1 specifies if the position of the token is even or odd, and $d_{model}$ represents the embedding size, which for the current research is equal to 256.

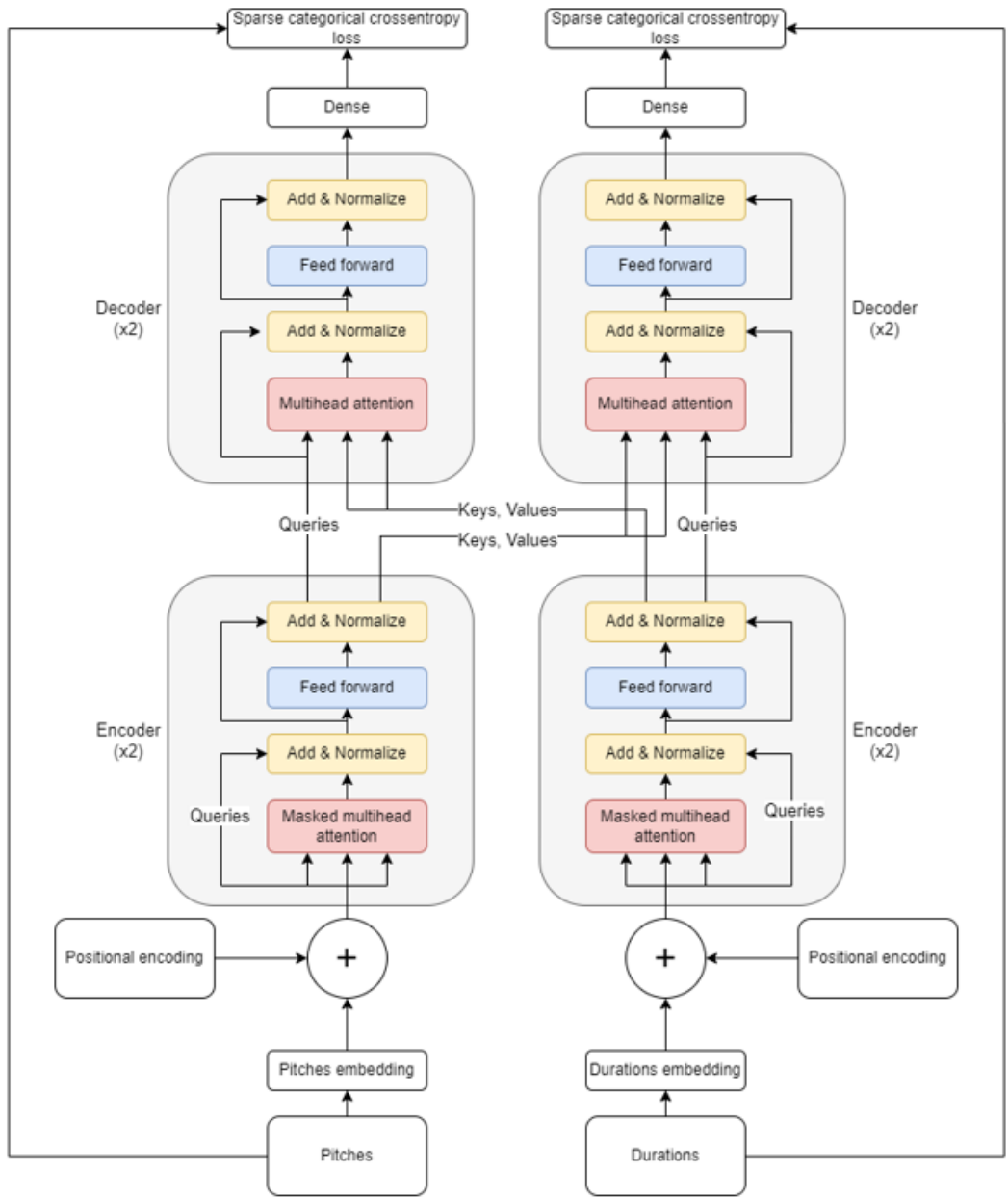

Fig. 2 - Modified Transformer architecture

The resulting token and position embedding combination is then sent as input into a masked multi-head attention layer. The multi-head attention is made up of eight separate attention mechanism layers (also called heads) that each accept a different part of the original tensor as input. Because the input from the multi-head attention layer has the size of 256 values for its last dimension (the first dimension being the batch size and the second dimension being the sequence length), each head receives a sequence of 150 values, and each value is embedded into a tensor of 32 values (256 divided by 8). Each head consists of a scaled dot product attention mechanism, where the mask is also applied for specific circumstances. At the end, the results from all the 8 heads are combined by concatenation in order to obtain the original embedding of 256 in size.

For both encoders a mask is used on the incoming data. This is required because the transformer neural network would otherwise learn that the input and output are the same. The transformer is required to act as a language model, which means that for each note and duration that it receives as input, it will attempt to create it as output by solely utilizing the notes and durations that came before it in the sequence.

The traditional vision transformer [10] concatenates each line of pixels from an image into one single list, and the mask is then applied to cover the pixels that must be generated (the target data). Because the pixels that are generated represent the output data, that means that the rest of the pixels from the image represent the input data. For the visual transformer neural network used in this article, instead of multiple line of pixels from an image, there are multiple voices from a melody. Each voice consists of notes that have a corresponding pitch and duration, so two separate sequences of pitches and durations can be made only for one voice. When the melody has multiple voices, then the concatenated sequences of pitches will become the input for the first encoder, and the concatenated sequences of notes will become the input for the second encoder.

In each encoder and decoder, the queries represent the input for the multi head attention layer and for the Add & Normalize layer. In the Add & Normalize layer, the input queries are added with the output from the multi head attention. The result is then regularized to avoid the entire network becoming overfitted. The regularization is made using a dropout layer with a dropout rate of 0.1 followed by a Layer Normalization layer. A feed forward neural network made up of two layers is utilized after the regularization. The first layer is a Dense layer with a Relu activation made up of 256 neurons (the size of the embedding), while the second layer is a Dense layer lacking the Relu activation previously used. The output of the feed forward neural network is then regularized in the same way that was used after the multi head attention layer.

At this point, the difference between a standard transformer and the design used in this study may be noticed. There is just one encoder and one decoder in a standard transformer. The original transformer's decoder is larger than the decoder used in this study since it performs the same functions as the encoder and decoder on the right side of the current design shown in Figure 2. In the current implementation, after each encoder another decoder will follow, so the architecture will look symmetrical.

The encoders and decoders are connected in the same way as they were in the original transformer architecture. The queries generated by the notes encoder are combined with the keys and values generated by the durations encoder, thus creating the input for the notes decoder, and the queries generated by the durations encoder are combined with the keys and values generated by the notes encoder creating the input of the durations decoder. Both the durations decoder and notes decoder are built in the same way as the encoders, with the exception that the mask is no longer utilized. Each decoder, such as the encoders, is made up of two successive decoders that increase the depth of the neural network.

Finally, for each of the decoders, a Dense layer is used to normalize the output data such that it matches the size of the original notes and duration embeddings. As a consequence, the Dense layer after the notes decoder has 126 neurons, whereas the Dense layer after the durations decoder has 833 neurons, resulting in the same original embedding.

The sparse categorical crossentropy was utilized to calculate the loss. Because the transformer's input data is identical to the target data (sequences of 50 tokens in size), but the neural network's output data is embedded with the original vocabulary's size (50, 126) and (50, 833), the sparse categorical crossentropy is better suited than the categorical crossentropy because it already implements the hot encoding of the target data, allowing the loss between the target data and the output data to be computed.

### 3.2. The second and third set of melodies

The second set of melodies is generated almost the same as it was in previous published research [19]. In that research, the durations of notes from the melodies were generated by the same method used by the MusicXML Creator platform [18] and the pitches for the same notes were generated using a sequence-to-sequence neural network made of GRU. In the current experiment, instead of a sequence-to-sequence neural network made of GRU. In the current experiment, instead of a sequence-to-sequence GRU, the original transformer architecture from Vaswani paper [5] is used. An advantage of the transformer architecture is that in the current research a much larger training corpus is used, and the transformer neural network is more suited because of its improved parallelization.

To train the transformer, the following hyperparameters were used: an embedding size of 128, eight heads multi-head attention and the decoder and the encoder were stacked 2 times. A sparse categorical cross-entropy loss was used for computing the error, and an Adam optimizer with a learning rate of 0.0001 was used for training the network. The input data is encoded using sine and cosine positional encoding. The neural network was trained for a total of 10 epochs on the training corpus. The training corpus consists of 90% of the data from the Lakh Midi dataset and the rest of 10% represents the testing corpus, which was used for computing the accuracy.

The same transformer was used in the generation of pitches for the third set of melodies, but the durations were generated using the Schillinger theory of rhythm rather than the sonification technique that was previously employed. The three melodies that belong to the third set were composed with the following rhythms:

- For melody 1, the duration sequences (3/8 1/4 1/8), (1/4 1/8 3/8), and (1/8 3/8 ¼) were labeled as A, B, and C. The rhythm for the melody is: A B C C A B for the first voice, B C A B B A for the second voice, and C A B A C C for the third voice.

- For melody 2, (3/8 1/8 1/8 3/8), (1 - full note), and (1/8 3/8 3/8 1/8) were labeled as C. The rhythm for the melody is: A B for the first voice, C A for the second voice, and B C for the third voice.

- For melody 3, (1/4), and (1/8 1/8 1/8) were labeled as A and B. The rhythm for the melody is: A B A A B A A A B B A A for the first voice, A B A A A B B A A A B A for the second voice, and A B A B A A A B A A A B A for the third voice. The duration sequences can further be labeled as follows: (A B A), (A A B), (B A A) being labeled as C, D and E. The rhythm for the melody can now be noted as: C C D E for the first voice, C D E C for the second voice, and C E C C for the third voice.

### 3.3.  The fourth set of melodies

The GPT3 Curie transformer [13] neural network was used to generate the fourth set of melodies.  The transformer was fine-tuned using 400 melodies that were randomly selected from Lakh Midi Dataset. The neural network produced an entire song using as input only the first 10 notes of a melody that was different from the ones found in set 1 but was produced by the same modified transformer that created the set.  This implies that GPT3 was used to continue the first notes of a melody that was also generated by a transformer.

Before training, the 400 melodies that were used for fine-tuning GPT3, were first converted to ABC format[4]. The reason for this decision was that ABC format encodes the notes in a much smaller text format than MusicXML[5]. This is important, because GPT3 Curie transformer is limited to an input of maximum 1000 tokens. If MusicXML encoding was used, the 1000 tokens would encode almost 90 consecutive notes. The same 90 consecutive notes can be encoded in ABC format using only 27 tokens. Other studies have also used with success ABC format for fine-tuning GPT3 [23].

---

[4] abc | home, abcnotation.com. https://abcnotation.com/
[5] musicXML for Exchanging Digital Sheet Music, MusicXML. https://www.musicxml.com/

The melodies have also been preprocessed. At this step, the newlines have been converted to "$". If the dollar sign is introduced as a symbol for defining a newline in the input data, the GPT3 model will recognize the pattern and utilize the symbol itself.

For training GPT3, input data (the first 10 notes of a melody) and target data (the rest of notes from the melody) must be paired in a JSON file. The music21[6] python library was utilized for this phase. The neural network's fine-tuning was done over a period of five epochs.

## 4. Evaluation and results

In order to evaluate the chosen music generation methods, a study was conducted in which 108 students (80 male and 28 female) from University Politehnica of Bucharest, Romania participated. The study was carried out with the help of a Google Forms questionnaire where the students were asked at the beginning to listen carefully to the 12 songs, having the possibility to listen to them again later. Each song lasted approximately 30 seconds, and this interval was chosen to allow students to complete the questionnaire in a short time interval. Then they were asked what grade (between 1 and 5) they would give for each individual melody (1 meaning that they don't like the respective song, and 5 meaning that they like the song very much). Also, the students were asked for what type of application the song could be used as a background: for a movie, for a game, for a creativity stimulation application or during an internet search.

To present the results, the arithmetic mean of the 108 grades given by the students for each individual song was calculated. Together with the arithmetic mean, the variance was also computed. The results are presented in Table 1, and Figure 3 shows the results of choosing the type of application in whose background the songs could be used.

---

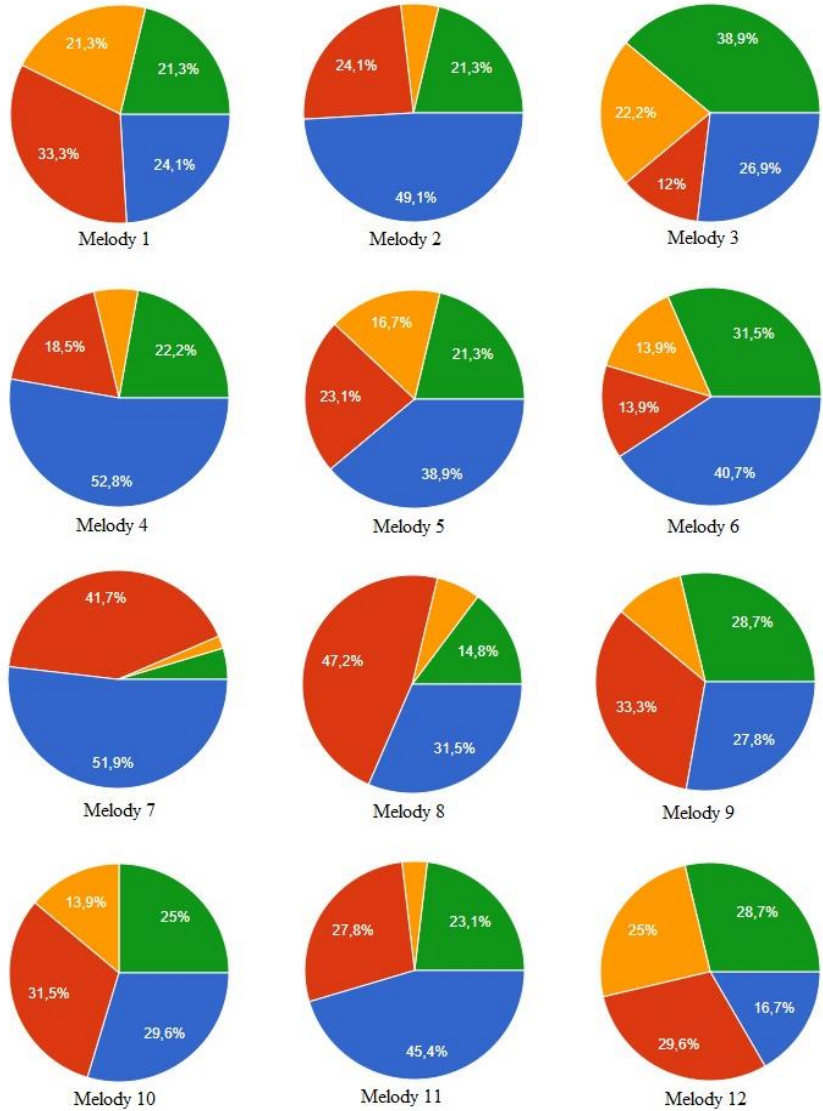

Fig. 3 - Results obtained from a study where the participants were asked to specify to what type of application 12 melodies could be used for a background music: for a movie (red), for a game (blue), for a creativity stimulation application (green) or during an internet search (yellow)

The results presented in Figure 3 were evaluated using the Chi-square test [24] (Chi statistic = 30.6, p < 0.05), followed by the Chochran Q test [25] for each type of application: movie (Q = 63.4, p < 0.01), game (Q = 74.9, p < 0.01), internet search (Q = 65.8, p < 0.01), and creativity stimulation application (Q = 49.2, p < 0.01). Both of these tests have revealed significant differences in the applicability of each method of music generation.

**Scores obtained through a study in which 108 students participated.**

| 1st set of melodies | Melody 1 | Melody 2 | Melody 3 |
|---|---|---|---|
| Scores mean | 2.65 | 3.44 | 3.31 |
| Variance | 2.00 | 1.33 | 1.35 |
| Standard deviation | 1.4 | 1.14 | 1.15 |
| 2nd set of melodies | Melody 4 | Melody 5 | Melody 6 |
| Scores mean | 3.09 | 3.15 | 2.06 |
| Variance | 1.65 | 1.49 | 0.84 |
| Standard deviation | 1.28 | 1.21 | 0.91 |
| 3rd set of melodies | Melody 7 | Melody 8 | Melody 9 |
| Scores mean | 3.38 | 2.57 | 3.43 |
| Variance | 1.20 | 1.44 | 1.25 |
| Standard deviation | 1.09 | 1.19 | 1.11 |
| 4th set of melodies | Melody 10 | Melody 11 | Melody 12 |
| Scores mean | 2.35 | 3.10 | 3.62 |
| Variance | 1.35 | 1.66 | 1.17 |
| Standard deviation | 1.15 | 1.28 | 1.07 |

For the aesthetic scores from which the results from Table 1 were obtained we calculated the following statistics:

- For each respondent, an average score per method was calculated by averaging the scores given for each song belonging to that method.
- The Shapiro-Wilk test was used to assess whether the average scores calculated for each method exhibit a normal distribution. It was found that for method 1 and method 2, the scores were not normally distributed (p. <0.05) so it was decided to use the Friedman test [26] (which is similar to ANOVA but does not assume that the variables are normally distributed).
- The Friedman test [26] was performed (with the R rstatix package) and resulted in significant differences between methods (p. < 0.01), but with a small effect size (Kendall = 0.08).
- Post-hoc Wilcoxon [26] pairwise tests were performed to see which methods exactly differed. Significant differences were found between Method 1 and Method 4 (p. <0.05), between Method 2 and Method 3 (p. <0.05), between Method 2 and Method 4 (p. <0.01) and Method 3 and Method 4 (p. <0.05).

In conclusion, the results indicate that method 4 was the most preferred among the four artificial intelligence algorithms for music generation. Method 3 and method 1 received comparable levels of preference, while method 2 was the least preferred. However, it is essential to note that this is not a definitive conclusion due to the small Kendall effect size.

## 5. Conclusions

The results obtained in the research presented in this paper indicate a significant difference between the four music generation methods that were selected for analysis. These findings provide compelling evidence that each approach has unique characteristics and outcomes in terms of music generation. The findings from this research highlight distinctions in students' perceptions regarding the suitability of each song as background music for specific applications. Notably the melodies generated by combining a neural network with Schillinger's theory of rhythm had a tendency to fit as background music to a movie. The research further revealed strong indications that Schillinger's theory of rhythm is significantly enhancing the aesthetic quality of music when compared to other sonification methods employed in this study. This result attests to the uniqueness of this music generation method, and they serve as a cornerstone for our future work. Our research will build upon this foundation, exploring the integration of Schillinger's theory of rhythm with diverse music generation techniques.

### Acknowledgements

The results presented in this article have been funded by the Ministry of Investments and European Projects through the Human Capital Sectoral Operational Program 2014–2020, Contract no. 62461/03.06.2022, SMIS code 153735.

## R E F E R E N C E S